\pdfoutput=1

\documentclass[11pt]{article}

\usepackage[final]{acl}

\usepackage{times}
\usepackage{latexsym}

\usepackage[T1]{fontenc}

\usepackage[utf8]{inputenc}

\usepackage{microtype}

\usepackage{amsmath}
\usepackage{multirow}

\usepackage{xcolor}

\usepackage{graphicx}
\DeclareGraphicsExtensions{.pdf,.png}
\graphicspath{ {figures/} }

\title{How well can Transformer-based semantic product search models generalize?}

\title{Towards Generalizable Semantic Product Search by Text Similarity Pre-training on Search Click Logs}

\author{Zheng Liu, Wei Zhang, Yan Chen, Weiyi Sun, Tianchuan Du, Benjamin Schroeder \\
  Wayfair LLC \\
  4 Copley Place, Boston, MA USA 02116 \\
  \texttt{\{zliu2,wzhang5,ychen4,wsun1,tdu1,beschroeder\}@wayfair.com} \\
}

\begin{document}
\maketitle
\begin{abstract}
Recently, semantic search has been successfully applied to e-commerce product search and the learned semantic space(s) for query and product encoding are expected to generalize to unseen queries or products. Yet, whether generalization can conveniently emerge has not been thoroughly studied in the domain thus far. In this paper, we examine several general-domain and domain-specific pre-trained Roberta variants and discover that general-domain fine-tuning does not help generalization, which aligns with the discovery of prior art. Proper domain-specific fine-tuning with clickstream data can lead to better model generalization, based on a bucketed analysis of a publicly available manual annotated query-product pair data.
\end{abstract}

\section{Introduction}
Semantic Search has been widely studied recently as a promising competitor to token-based inverted index search that had dominated information retrieval (IR) and question answering (QA) for decades.
However, in e-commerce product search, the quality of the learned embedding space is not well understood, especially on the quality of generalization to unseen queries, unseen products, or even unseen query-product pairs. 
The generalization is particularly interesting and important because in Wayfair 80\% of queries are new in the monthly basis, and new products are added everyday. 
In this paper, we will investigate the generalization issue by seeking answers to the two following questions: 
\begin{enumerate}
    \item How well does the learned space generalize to unseen queries/products/query-product combinations? 
    \item Are we able to improve generalization by pre-training, and how?
\end{enumerate}    

Arguably, some pre-trained models, especially the ones pre-trained with domain-specific clickstream data, may outperform the others on generalization. To validate the hypothesis, we train two-tower semantic search models using three different categories of base models for comparison: not pre-trained (Roberta-base \cite{liu2019roberta}, Roberta-large), general-domain pre-trained with text similarity (SimCSE \cite{gao-etal-2021-simcse}) and domain-specific pre-trained with text similarity (SimCSE that are continuously pre-trained with clickstream data). 

To observe how the models perform, we take a publicly available product search dataset \cite{wands} and partition the data into buckets according to whether the query/product/query-product pairs have been visited during training. By comparing the performance of the above models, we observe that query representations may have been insufficiently learned during two-tower semantic search fine-tuning as opposed to product representations, and the model variant which is pre-trained on query-query text similarity has a visible impact on models' generalization power.

This paper is organized as follows: 1) we first introduce related work, then 2) explain the proposed pre-training approach, and finally 3) discuss results and analysis. 

\section{Related Work}

\paragraph{Semantic search in product search}
Recently, semantic search has been explored in e-commerce product search \cite{nigam2019semantic, huang2020embedding, zhang2020towards, li2021embedding}, in which queries and products are embedded as vectors through representation learning and search is then conducted in the vector space. These implementations generally follow a two-tower approach: one group of stacked neural network layers, usually referred to as ``tower'', processes the query, and another tower processes the product. This approach is favored in production due to its ability of decoupling query and product embedding process, so that products can be embedded offline and search can be accelerated in real-time services. In search systems, the semantic search is usually used as candidate generators focusing on core-relevance, and often followed by dedicated ranking layers that capture other e-commerce attributes such as price, popularity, quality, etc. The two tower model of this work is described in Section 3.5.

\paragraph{Generalization for two-tower neural models} Deep Passage retrieval has become a popular approach recently for IR. In Question Answering, Lewis et al. \shortcite{lewis2021question} observed that when using the two-tower approach for relevant passage retrieval stage, deep learning models tend to memorize training instances and perform substantially worse on unseen questions or unseen answers. They argued the generalization on questions is the major issue, which is similar to our observation in product search. Mao et al. \shortcite{mao-etal-2021-generation} tried to address the issue by explicitly retrieving question-related texts as query expansions and showed improvements on model generalization to unseen question/answers, yet we take the route of query similarity learning by using automatically mined query pairs from product co-click behaviors. 

\paragraph{Pre-training for Neural Retrieval} Chang et al. \shortcite{chang2020pretrain} pre-trained neural retrieval-based QA models with carefully designed synthetic tasks, followed by Guu et al. \shortcite{guu-2020} with a knowledge-augmented training task and Sachan et al. \shortcite{sachan-etal-2021-end} with combining ICT and masked salient spans training. O{\u{g}}uz et al. \shortcite{ouguz2021domain} instead enhanced training with domain-matching Reddit post comment pairs to pre-train a dense retrieval model and also observed positive influences to open domain QA. Differently, our approach explores the clickstream graph structure instead of mining additional data.

\paragraph{Text Representations Learning} Our approach closely relates to the graph node embeddings within texts \cite{mikolov2013distributed} or the linked graphs \cite{hamilton2017inductive} rather than node embeddings that involve node relation types \cite{bordes2013translating} or graphical neural networks that aggregate neighbouring nodes in a parametric fashion \cite{li2016gated,kipf2016semi}. Recently, noisy contrastive learning has enjoyed significant success \cite{gao-etal-2021-simcse} for regularizing text encoders, and shows that such encoders can effectively serve as a starting point for downstream task fine-tuning. In retrieval tasks, Gao and Callan \shortcite{gao&callan2021} pre-trained text encoders with synthetic passage similarity tasks using noisy contrastive loss, assuming that in Wikipedia two adjacent passages in the same article are similar in semantics rather than those from two different articles. In contrast, in product search, we assume two queries leading to the same product browse, or two products purchased under the same query, are similar.

\begin{table*}[t]
    \centering
    \begin{tabular}{c|c|c|c}
        \textbf{first query} & \textbf{second query} & \textbf{relation} & \textbf{score}  \\
        \hline\hline
girls study desk&outdoor shed&Entirely different intents&0\\
bathroom window curtains short & curtains bedroom &Similar (not entail/co-intent) &1\\
52 inch ceiling fan with light & Ceiling fans with lights & Entailment & 2 \\
Fishing storage & Fishing tackle storage & Same intent & 3 \\
    \end{tabular}
    \caption{Query similarity manual annotation samples. Note that first and second query is order-sensitive. However, as similarity strength, we disregard directionality.}
    \label{tab:qqdata}
\end{table*}

\section{Approach}
This section describes how models are pre-trained, fine-tuned, and evaluated on clickstream data. Section 3.1 sets up preliminaries and math notations for the product search problem; Section 3.2, 3.3 and 3.4 describe three different pre-training approaches; Section 3.5 discusses how the model is fine-tuned and evaluated in the product search problem.
\subsection{Preliminaries}
A click graph is denoted as $\mathcal{G}=\{{V},{E}\}$ where $V$ is the set of nodes either being a query $q$ or a product $p$ which is the concatenation of title, brand name, and class name \footnote{A product can have multiple class names with hierarchical structure (e.g. class ``sofa'' belongs to class ``living room furniture''). Here we used the high level product class name (e.g. ``living room furniture''). }. An edge $E$ connecting $q$ and $p$ suggests a customer showed interest to product $p$ when searching with query $q$.

The core of semantic search is to learn query and product representations. Intuitively, a good representation that generalizes may put similar concepts ($p$ or $q$) closer in the semantic space, while dissimilar ones further apart. 

There are three node type combinations $(p,p)$, $(q,q)$, and $(p,q)$, the last of which is the goal of semantic search. Arguably, learning solely on $(p,q)$ may cause generalization issue when $(p,p)$ and $(q,q)$ relations can not be implicitly learned from $(p,q)$ objective. To verify if this hypothesis holds in our semantic search problem, we design three pre-training tasks to warm up Transformers: the $(q,q)$ pre-training task, the $(p,p)$ pre-training task, and the $(q,q)$ and $(p,p)$ pre-training task. Afte pre-training, we fine-tune the model on the $(p,q)$ learning objective. In the remaining of this section, we describe the three pre-training and the fine-tuning approaches.

\subsection{Query Similarity Learning}\label{sec:qq}
\paragraph{qq Model} If two queries $q_1$ and $q_2$ link to the same product $p$ in graph $\mathcal{G}$, we regard  $(q_1, q_2)$ as one co-clicked query pair. After collecting the pairs, we use them to fine-tune a SimCSE-large \cite{gao-etal-2021-simcse} model to obtain a model we call $\textbf{qq}$ model.
\paragraph{Pair Sampling} We follow the 3 steps below to mine pairs:
1) sample $q_1$ by $P(q)$, the query distribution obtained from data;
2) for $q_1$, we collect top-K popular products $p_1, ... p_K$ by frequency, and uniformly sample one product $p$ from them \footnote{``Uniformly'' as a reasonably simple yet arguably effective approach to mitigate exposure bias without complex user behavior modeling}, and 
3) at last, for $p$, we collect all its related queries and sample one query as $q_2$ uniformly.
We repeat this process $N$ times to obtain $N$ pairs where $N$ is set to 1 million.

\paragraph{Training}
We train the model with a noisy contrastive learning objective as in \cite{gao-etal-2021-simcse} with default settings. We train the model with 3 epochs. We fine-tune on top of the unsupervised SimCSE-large model and observe the additive effect of open-domain and domain-specific learning.

\paragraph{Similarity Evaluation}
To evaluate the query representation, we invite 3 domain experts to annotate the degree of query pairs into 4 relation types: not related (0), similar but not entail (1), entailment (2) and identical (3), where the number represents the semantic strengths, the higher the more related they are. A total of 1056 random query pairs are extracted from click sessions and the relation types are evenly distributed. The inter-annotator agreement was 90\% after 2 rounds of annotation and curation. A sample of the annotation data is shown in Table \ref{tab:qqdata}.

We calculate vector cosine between two queries as similarity score, and compare it to the ground-truth by Spearman Rank Correlation with significant tests. The result is shown in Table \ref{tab:qqtable}. As we can see, the model fine-tuned with co-click data achieves the best performance. Surprisingly, ``Ours-unsup'' learning noisy contrastive loss by masking a single query twice and learning their similarities (similar to SimCSE-unsup with open-domain texts), does not lead to a better performance on query pairs, suggesting that the key of query similarity learning is to capture rich domain-specific text similarity knowledge rather than text robustness and regularity in semantic space. This message encourages us to further extend it to product similarity learning (this time, without manual evaluation).

\begin{table}[t]
    \centering
    \begin{tabular}{l|l}
         \textbf{Model} & \textbf{SP} (P-value) \\
         \hline\hline
         Roberta-large & 0.287 (1.8e-21) \\
         SimCSE-large-unsup& 0.467 (2.9 e-58)\\
         SimCSE-large-sup & 0.426 (7.9e-48) \\
         \hline
         Ours-unsup & 0.458 (2.2e-48)\\
         Ours-sup (co-click) & 0.546 (5.8e-82) \\
    \end{tabular}
    \caption{Comparison of different Models's Spearman Rank Correlation on representation evaluation data. SimCSE--large-unsup is the representation learned \textit{unsupervisedly} with noisy contrastive loss; SimCSE-large-sup is subsequently learned from \textit{supervised} text entailment data. In our settings, the definition of unsup and sup changed a bit: Ours-unsup is the model learned with unsupervised objective (masked language modeling) with noisy contrastive loss, where Ours-sup is learned from query co-click data described in section \ref{sec:qq}. }
    \label{tab:qqtable}
\end{table}

\subsection{Product Similarity Learning}

\paragraph{pp Model} The product representation learning is performed similarly to the query representation learning. We start from SimCSE-large-unsup model, and fine-tune $(p,p)$ pairs that are obtained from the process similar to the sampling approach in section \ref{sec:qq} with the role of query and product swapped. The created 1 million pairs are used to fine-tune the SimCSE model and obtain a model dubbed \textbf{pp} model. Different from \textbf{qq} model, we did not obtain manual annotation data for \textbf{pp}, thus we let the model train 2 epochs until convergence.

\subsection{A Combined Model}
\paragraph{$\textbf{qq+pp}$ Model} In this approach, we sequentially pre-trained the model with both $(q,q)$ data and $(p,p)$ data, to examine the additive effect where a model is fine-tuned with both query similarities and product similarities.
We take the $\textbf{qq}$ model and continue to pretrain it with the above $(p,p)$ data until convergence, and obtain a model called $\textbf{qq+pp}$. Indeed, we may risk catastrophic forgetting since arguably the $(p,p)$ pairs may overwrite the captured $(q,q)$ information. Although we have observed such forgetting to happen, its effect is mild at best. 

\subsection{Fine-tuning for Semantic Search}

\paragraph{pq Learning} After pre-training, the semantic search model was fine-tuned on product search tasks: given a search query, the task is to retrieve a list of relevant products from the product catalog. Specifically, both queries and products were represented as vectors in the same latent space, and products were retrieved according to cosine similarities w.r.t. the queries. 

\begin{figure}[h]
    \centering
    \includegraphics[width=8.7cm]{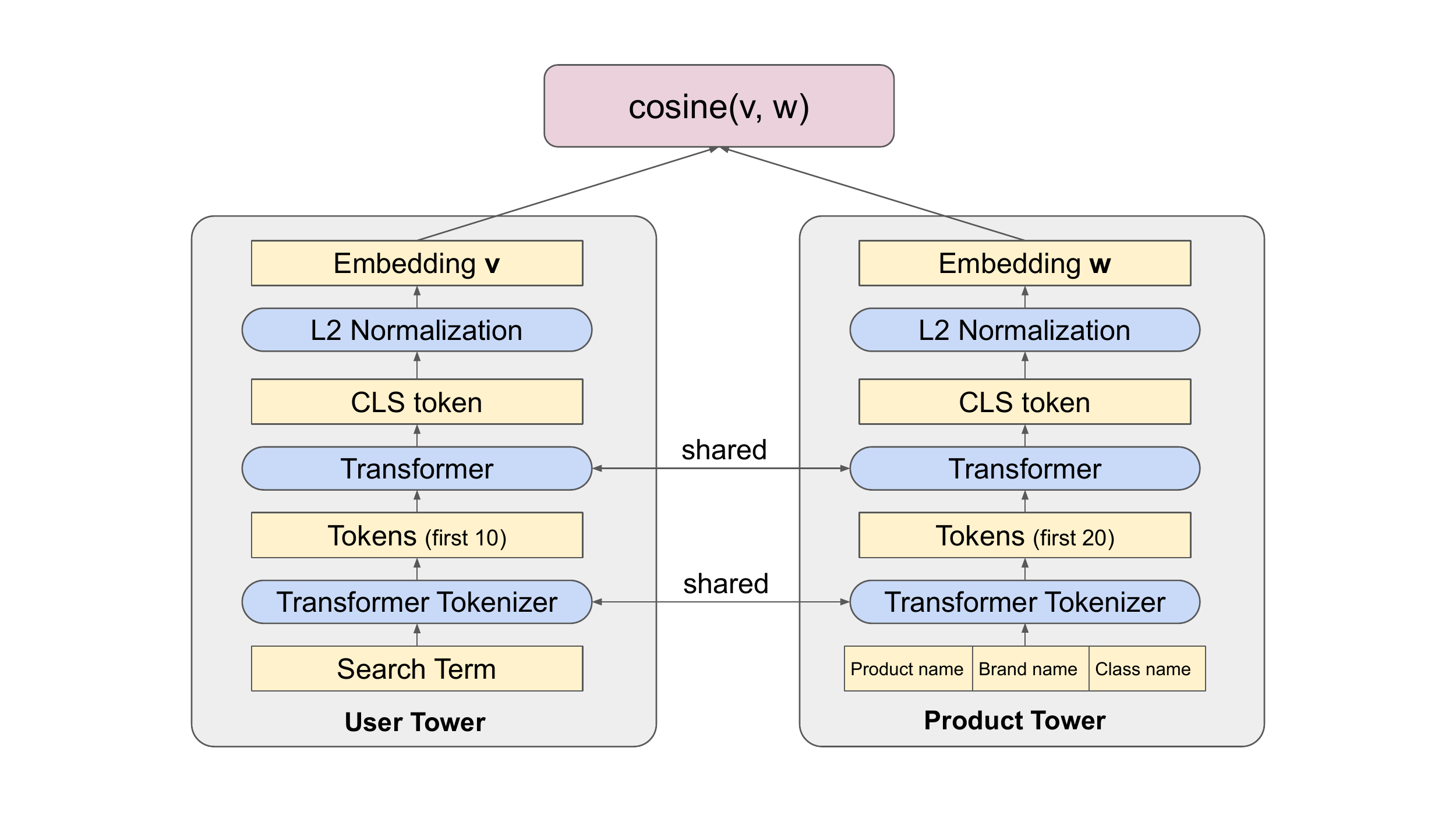}
    \caption{Structure of semantic search models on product search tasks}
    \label{fig: model_structure}
\end{figure}

As shown in figure \ref{fig: model_structure}, we adopted a two tower approach to fine-tune the semantic search model: a user tower embeds search queries into vectors, while a product tower embeds product information (e.g. product name, product brand name, product class name) into vectors. The two towers share the same tokenizer and transformer model. We took the last hidden layer's CLS token representation and L2-normalized it as the final output embedding vector. Then, the relevance score was calculated as the cosine similarity between search query embedding and product embedding vectors.

To train the model, we adopted the contrastive framework and utilized the cross-entropy objective with in-batch negatives \cite{chen2020simple}. In each mini-batch, we sampled $n$ semantically relevant (query, product) pairs $D=\{(q_i, p_i)\}_{i=1}^n$, and calculated the training objective for the i'th pair $(q_i, p_i)$ as:
\begin{equation}
    l_i = - \frac{cos(f(q_i),g(p_i))}{\tau} + log \sum_{j=1}^{n}e^{ \frac{cos(f(q_i),g(p_j))}{\tau}}
\end{equation}
$f$, $g$ represent the user and product tower respectively. $\tau$ is a temperature hyperparameter. All layers of Transformer models are fine-tuned in the product search task.

\paragraph{Data}
In the experiment, $(q,p)$ pairs are collected from click graph $\mathcal{G}$. As show in table \ref{tab: q_s_pairs}, each search query and its resulted add-to-cart products (if any) are constructed as semantically relevant (query, product) pairs. The training dataset contains 31M pairs from 2019 and 2020 records of a e-commerce search engine, and the dev dataset contains 100k pairs randomly sampled from 2021. We evaluated model's performance on the \emph{WANDS} dataset \cite{wands}, which is a human annotated dataset for evaluating product search relevance. It contains (query, product) pairs with relevance labels annotated by humans in three different levels: relevant, partially relevant, and irrelevant. Table \ref{tab: product_search_dataset} shows details about the training, dev and test dataset.

\begin{table*}[t]
    \centering
    \begin{tabular}{l|l}
        \textbf{query} & \textbf{product}  \\
        \hline\hline
        rose gold wing chair & alyka adonis wingback chair, red barrel studior, accent chairs \\
        led lights strip lights & doynton strip light, the holiday aisle, under cabinet lighting \\
        sun shelter & messina aluminum patio gazebo, sojag, canopies \& gazebos \\
    \end{tabular}
    \caption{Example of semantically relevant (query, product) pairs}
    \label{tab: q_s_pairs}
\end{table*}

\begin{table}
    \centering
    \begin{tabular}{l|c|c|c}
        & train & dev & test \\
        \hline\hline
         total pairs & 32M & 100K & 234K \\
         unique queries & 9.5M & 76K & 483\\
         unique products & 2.1M & 88K & 43K\\
         unique pairs & 21M & 99K & 234K \\
    \end{tabular}
    \caption{Dataset summary. Train and dev datasets are collected from historical browsing contexts; the test dataset (WANDS) is a human annotated dataset for evaluating product search relevance \cite{wands}. }
    \label{tab: product_search_dataset}
\end{table}

\paragraph{Training and Evaluation}
6 different Roberta variations are fine-tuned in this experiment. They are Roberta base, Roberta large \cite{liu2019roberta}, unsupervised SimCSE based on Roberta large \cite{gao-etal-2021-simcse},  and three in-house pretrained models based on the SimCSE model using query-query, product-product, and query-query plus product-product similarity data respectively.

All of the models have the same training setup. For each model, we ran 4 epochs on the training dataset with mini-batch size of 256. Model checkpoints were saved every 6000 mini-batches, for the purpose of model selection. Adam optimizer with learning rate of $1e^{-5}$ was used to train the model. The temperature hyperparameter was set as 0.07.

In model selection, for each model we select the best performing checkpoint on the dev dataset to be the final trained model. The evaluation metric is the top-k retrieval accuracy, defined as the ratio of \emph{(query, product)} pairs for which the top-k retrieved products of the \emph{query} contains the corresponding \emph{product}. The retrieval space for dev dataset is the unique products in dev dataset.

To evaluate the final trained model, we calculated nDCG@k scores on the WANDS dataset. The ground truth relevance score is derived from human annotation labels: relevant is 1, partial relevant is 0.5, and irrelevant is 0.

For both the retrieval accuracy and nDCG scores, we set $k=50$, which roughly equals to the number of products shown on the first page of search results.

\section{Results}
We evaluated the models' performance on the WANDS dataset. As noted in Table \ref{tab: product_search_dataset}, this dataset contains 234k annotated (query, product) pairs. We further expand this dataset exhaustively to $20.6M$ pairs so that it contain every possible combination of unique query and unique product from the WANDS dataset. If a (query, product) pair is not annotated, we assume its relevance score is 0. Then, we assign these (query, product) pairs into different buckets, according to whether the query, the product, or the (query, product) pair exists in the training data or not. Details about bucket assignments are shown in Table \ref{tab: product_search_results}. Finally, for each bucket, we calculate the averaged per-query nDCG@50 score.

\begin{table*}[t]
    \centering
        \begin{tabular}{l||c|c|c|cccc}
        \hline\hline
        &&&&\multicolumn{4}{c}{Unseen sub-buckets} \\
        Bucket index          & 1 & 2 &3 & 4 &5 &6 &7 \\
        Bucket name           & Overall   & Seen      & Unseen    & q+, p+    & q+, p-    & q-, p+    & q-, p-\\ \hline
        Bucket ratio          & 100\%         & 0.03\%         & 99.9\%         & 29.87\%       & 31.60\%         & 18.72\%         & 19.79\% \\
        \hline\hline
        Roberta base                    & 0.7812        & 0.9335        & 0.7722    & \bf{0.7539}    & 0.7894    & 0.6962    & 0.7249\\
        Roberta large                   & 0.7830        & 0.9337        & 0.7730    & 0.7511    & 0.7893    & 0.6942*    & 0.7177\\
        SimCSE large                    & 0.7775*       & 0.9315        & 0.7679*    & 0.7491    & 0.7875    & 0.6884*    & 0.7179*\\
        Query-query (qq) pretrain       & \bf{0.7866}   & \bf{0.9351}   & \bf{0.7760}    & 0.7477*    & \bf{0.7918}    & \bf{0.7057}    & \bf{0.7276}\\
        Product-product (pp) pretrain   & 0.7802*       & \bf{0.9351}   & 0.7700*    & 0.7460*    & 0.7875    & 0.7003    & 0.7201\\
        qq+pp pretrain  & 0.7845        & 0.9319        & 0.7747        & 0.7489    & 0.7886    & 0.7006    & 0.7270\\
        \hline\hline
        \end{tabular}
        \caption{nDCG@50 of fine-tuned models on product search tasks, evaluated by the WANDS dataset. The model names indicate the base model before fine-tuning. The ``overal'' bucket contains all (query, product) pairs in the testing dataset, the ``seen'' bucket contains (query, product) pairs from the testing dataset that also exist in the training dataset, and the ``unseen'' bucket contains (query, product) pairs from testing dataset that does not exist in the training dataset. The ``unseen'' bucket is further split into four different sub-buckets, according to whether the specific query or product exists in the training dataset or not. ``q'' represents query, ``p'' represents product; ``+'' represents existing in the training dataset, ``-'' represents not existing in the training dataset. Bucket ratio represents the size of current bucket with respect to the entire test dataset.  In nDCG scores, ``*'' indicates that number is statistically smaller than that column's largest (bold-faced) number, tested by one-side t-test with 0.05 p-value. nDCG results at other k's are presented in the Appendix \ref{apd:ndcg}.}
    \label{tab: product_search_results}
\end{table*}

\paragraph{RQ1: How well does the learned semantic space generalize?} 
Table \ref{tab: product_search_results} shows each model's $nDCG@50$ scores on different buckets of the WANDS dataset. The "seen" bucket has much higher nDCG scores than the "unseen" bucket, indicating a large opportunity area of generalization for all models. By looking further into the amortized unseen sub-buckets, it shows that queries and products follow very different generalization behaviors. For example, the queries seen in the training data (q+, bucket 4, 5) usually performs better than the unseen queries (q-, bucket 6, 7). Yet counter-intuitively, the products seen in training under-perform the unseen ones (bucket 4 vs 5, 6 vs 7), which is against the common understanding where seen products may perform better than unseen ones. Arguably, this difference between query and products can be naively explained as products being relatively well learned compared to queries. We will revisit this point in analysis by graphical evidence. 

\paragraph{RQ2: Are we able to improve generalization by pretraining?}
In Table \ref{tab: product_search_dataset}, two domain-pretrained models (the qq model, and the qq+pp model) achieved top-two overall scores, showing the effectiveness of domain-specific pretraining approach. This, again, suggests the importance of query representation learning. Also, the SimCSE-large model performs worse than the Roberta-large model, suggesting that not all general-domain pretrainings are beneficial for domain-specific tasks, which may be due to the fact that the task of semantic search relies on asymmetric query-product relation encoding more than symmetric point-wise similarity. In contrast to SimCSE-large, the domain-specific query-query pretraining may have served the purpose of injecting domain knowledge mined from graph structure into text encoders, rather than a means for representation regularization, and such consistent improvement may have opportunity to be further improved given its moderate improvements over other models. Last, the Roberta-large model outperforms the Roberta-base model as expected due to a larger model capacity, yet the capacity for memorization is a draw between the two, based on the numbers in bucket 2.
 
\section{Analysis}
To understand why queries and products follow different generalization pattern in terms of nDCG scores observed in RQ1, we further analyzed the query-product cosine score distribution for different buckets. To make distribution comparison fair across buckets, in each bucket we only keep a subset of (query, product) pairs that are annotated as "partial relevance".
The model we used to generate the cosine score distributions is the SimCSE-large model that is first \textbf{qq}-pretrained and then \textbf{pq} fine-tuned on click graph data, the best model according to Table \ref{tab: product_search_results}.

\begin{figure}[t]
    \centering
    \includegraphics[width=8.5cm]{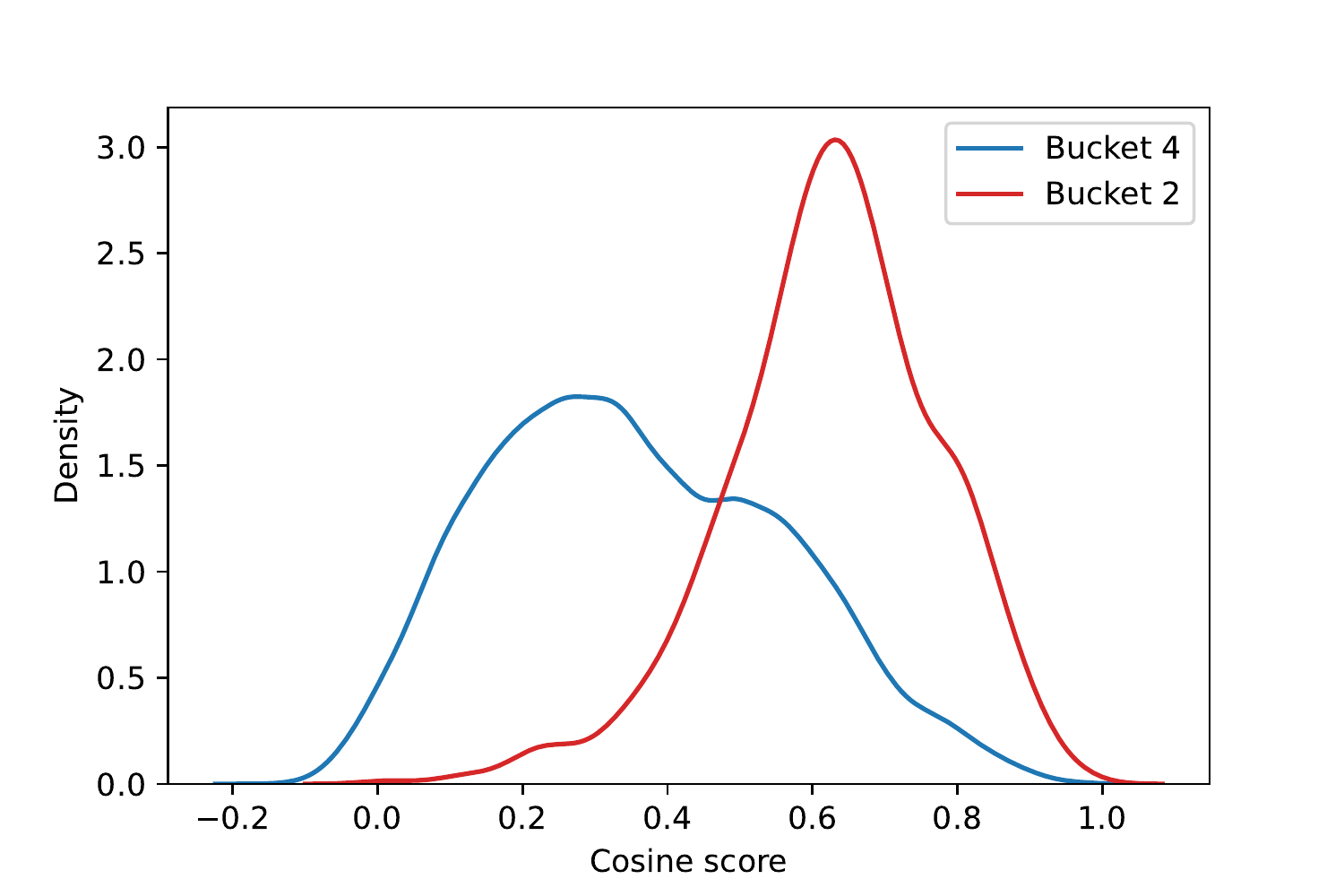}
    \caption{Query-product cosine score distribution for different buckets. The model is firstly pretrained by query-query similarity data, then fine-tuned in the product search task. Bucket 2 represent (query, product) pairs already seen by the model in training. In bucket 4, both query and product are seen, but the combination is not seen by the model in training. Bucket 2's mean cosine score (0.6253) is significantly higher than bucket 4 (0.3562).}
    \label{fig: cosine_score_42}
\end{figure}

First, the models perform better on seen query-product combinations than unseen combinations (the query and the product has been individually seen in training, but not the combination). The bucket 2 from Figure \ref{fig: cosine_score_42} shows a well learned cosine score distribution for seen query-product pairs, and the scores concentrate around a high similarity range around $0.65$. In contrast, for bucket 4, although queries and products are individually seen by the models during training, the combination is novel to them, and it increases the models' uncertainty, which is demonstrated by a significant similarity score spread-out accompanied by a shift to a lower median around $0.3$. In general, the figure shows the difficulty of generalization of deep learning-based semantic search models on even the simpler task of dealing with unseen combinations. This suggests the generalization may be beyond the discussion of learning representations for an individual query or product, and rather about the geometry of the semantic space to properly maintain the distance between seen and unseen texts after semantic search as asymmetric embedding learning that ``disturbs'' a well-defined text representation space to encode text rankings.

\begin{figure}[h]
    \centering
    \includegraphics[width=8.5cm]{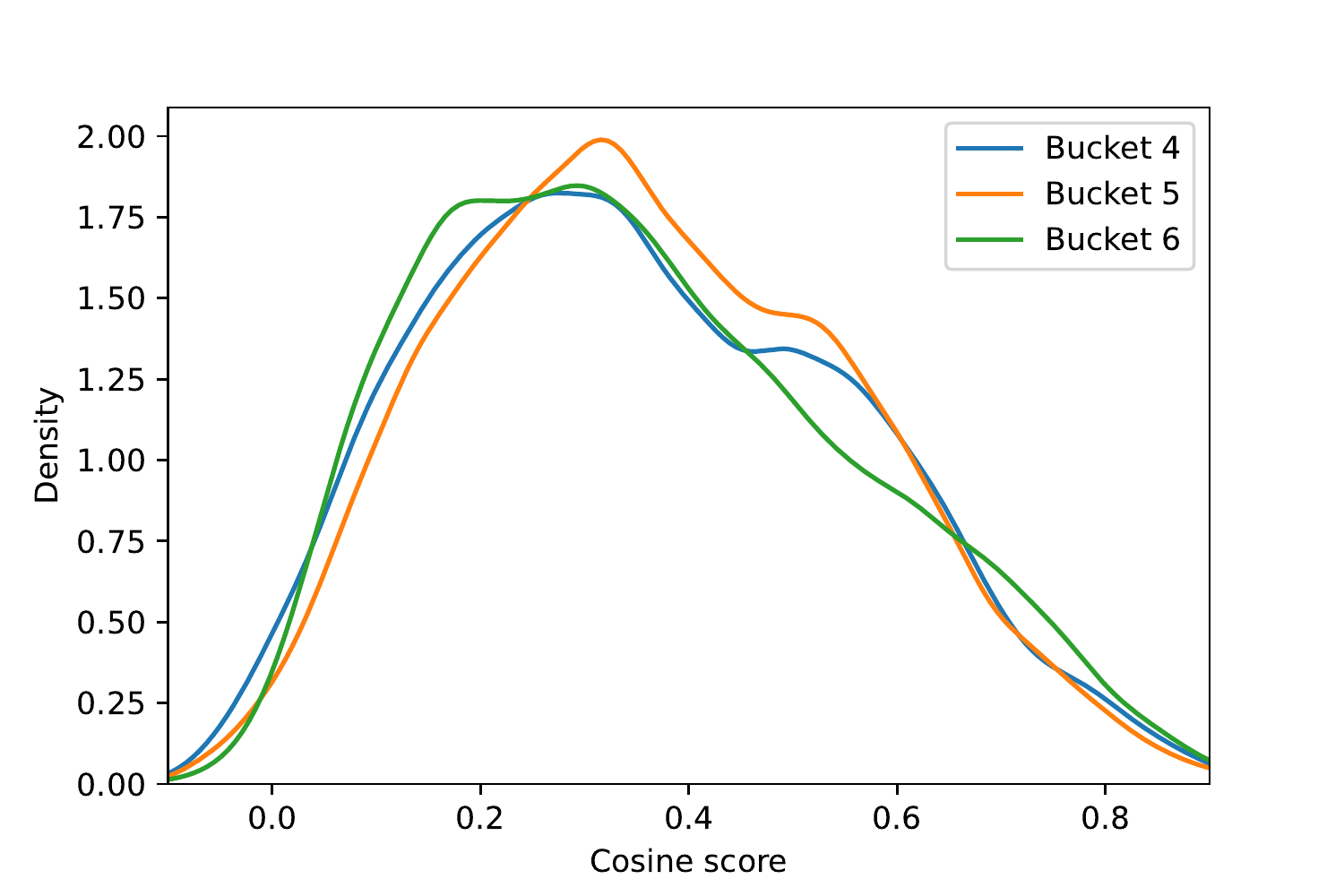}
    \caption{Query-product cosine score distribution for bucket 4, 5 and 6. Their mean cosine scores are 0.3562, 0.3655, and 0.3556 respectively. In bucket 4, both query and product are seen, but the combination is not seen by the model in training. In bucket 5, only the query is seen by the model in training. In bucket 6, only the product is seen by the model in training.}
    \label{fig: cosine_score_456}
\end{figure}

Second, as shown in Figure \ref{fig: cosine_score_456}, the visibility of queries and products in training data poses opposite impacts to a model's behavior. 

1) For seen products, comparing the curves of bucket 5 and 4, 1) counter-intuitively, we observe that 4 have a lower peak of around $0.25$ as opposed to that of bucket 5 of around $0.3$, which suggests that seen products have been ``attracted'' to seen queries in the training data, thus leading to a biased product representation. Yet, unseen products do not get to go through such an attraction process, and the representation may have maintained a neutral and generalize-able status in semantic space, leading to an overall higher and concentrated peak of 5 as opposed to a flat and lower peak of 4. This phenomenon might have suggested that a large semantic space may be over-parameterized that separately adjusting an individual product's representation without affecting its geometrically closer neighbours may have become easy. Thus, an un-generalizable learning in generalizable models might have happened. 

2) For seen queries, however, by comparing the cures of bucket 6 and 4 (the same as 7 and 5), we found that queries that have been seen in the training data lead to better semantic similarity with products in general than unseen queries, demonstrated by the distribution mass of the blue curve towards a higher similarity range than the green curve. The green line's peak being extended towards the lower cosine score range demonstrates that the representation of seen queries may have improved instead of overfit. In other words, this may also suggest that query representation may have not been fully learned. From Table \ref{tab: product_search_results} we can also observe a similar hint that query-query pre-trained model is the most effective at improving semantic search performance from product ranking's perspective. In stark contrast to seen products being overfit, seen queries gets to be learned and improved during semantic search's \textbf{pq} learning process.

By comparing the above two cases, we have seen that queries may have learned slower than products in semantic search. The reason may be due to the fact that queries are usually much richer and more diverse (we have tens of times more unique queries than unique products), requiring a text encoder to observe way more cases for a query to learn representation well; whereas products are relatively fewer, well-formed and limited in diversity in product search. However, for a company that has billions of products in inventory, this observation may not hold true. Yet, the observation indeed resonates with that of question answering and web search, which suggests learning generalizable semantic search through improving query representation learning may still be a promising direction to follow. 

\section{Conclusions and Future Work}
This paper examines the encoder generalization problem in the setting of semantic product search. We observe that query representation learning still is the bottle neck compared to the product representation learning under the semantic search training objective, the problem of which is surfaced by the superior performance of the domain-specific query pair pre-trained encoder model. We also surprisingly found that large, general-domain and unsupervisedly regularized Roberta variants such as SimCSE do not guarantee superior performance compared to Roberta-base or large, which may have suggested that semantic search is an \textit{asymmetric} relationship learning task rather than symmetric, similarity learning task, since the semantic space has to encode much more complex relations (e.g. node priorities in a list) beyond point-wise similarity. 

Although being a proof of concept, the proposed pre-training approach could start the conversation about how to properly pre-train encoders for semantic product search, and the results suggest that the knowledge hidden in the click graph can be better explored by defining proper training tasks, either transforming the task into a sequential task learning setting or a multi-task setting. In the future, a transfer learning setting that leverage richer domain knowledge and semantic structures from relevant sources such as class and class taxonomy of queries may be promising. Also, data augmentation to enhance query representation learning, or novel learning curriculum/objective/regularization that can help balance query and product learning in the semantic search objective may also be valuable for exploration.

\section*{Acknowledgements}
The authors would like to acknowledge previous and current members of the Wayfair Search and Recs team for full support on this research: Shujian Liu, Archi Mitra, John Castillo, Pooja Ranawade, John Murray, Avanti Patil, Zhilin Chen, Jeremy Myrland, and Matt Herman. We would also love to thank the suggestions and comments from reviewers to improve this paper.

\bibliography{anthology,custom}

\begin{thebibliography}{20}
\expandafter\ifx\csname natexlab\endcsname\relax\def\natexlab#1{#1}\fi

\bibitem[{Bordes et~al.(2013)Bordes, Usunier, Garcia-Duran, Weston, and
  Yakhnenko}]{bordes2013translating}
Antoine Bordes, Nicolas Usunier, Alberto Garcia-Duran, Jason Weston, and Oksana
  Yakhnenko. 2013.
\newblock Translating embeddings for modeling multi-relational data.
\newblock \emph{Advances in neural information processing systems}, 26.

\bibitem[{Chang et~al.(2020)Chang, Yu, Chang, Yang, and
  Kumar}]{chang2020pretrain}
Wei{-}Cheng Chang, Felix~X. Yu, Yin{-}Wen Chang, Yiming Yang, and Sanjiv Kumar.
  2020.
\newblock \href {http://arxiv.org/abs/2002.03932} {Pre-training tasks for
  embedding-based large-scale retrieval}.
\newblock \emph{CoRR}, abs/2002.03932.

\bibitem[{Chen et~al.(2020)Chen, Kornblith, Norouzi, and
  Hinton}]{chen2020simple}
Ting Chen, Simon Kornblith, Mohammad Norouzi, and Geoffrey Hinton. 2020.
\newblock A simple framework for contrastive learning of visual
  representations.
\newblock In \emph{International conference on machine learning}, pages
  1597--1607. PMLR.

\bibitem[{Chen et~al.(2022)Chen, Liu, Liu, Sun, Baltrunas, and
  Schroeder}]{wands}
Yan Chen, Shujian Liu, Zheng Liu, Weiyi Sun, Linas Baltrunas, and Benjamin
  Schroeder. 2022.
\newblock Wands: Dataset for product search relevance assessment.
\newblock In \emph{Proceedings of the 44th European Conference on Information
  Retrieval}.

\bibitem[{Gao and Callan(2021)}]{gao&callan2021}
Luyu Gao and Jamie Callan. 2021.
\newblock \href {http://arxiv.org/abs/2108.05540} {Unsupervised corpus aware
  language model pre-training for dense passage retrieval}.
\newblock \emph{CoRR}, abs/2108.05540.

\bibitem[{Gao et~al.(2021)Gao, Yao, and Chen}]{gao-etal-2021-simcse}
Tianyu Gao, Xingcheng Yao, and Danqi Chen. 2021.
\newblock \href {https://doi.org/10.18653/v1/2021.emnlp-main.552} {{S}im{CSE}:
  Simple contrastive learning of sentence embeddings}.
\newblock In \emph{Proceedings of the 2021 Conference on Empirical Methods in
  Natural Language Processing}, pages 6894--6910, Online and Punta Cana,
  Dominican Republic. Association for Computational Linguistics.

\bibitem[{Guu et~al.(2020)Guu, Lee, Tung, Pasupat, and Chang}]{guu-2020}
Kelvin Guu, Kenton Lee, Zora Tung, Panupong Pasupat, and Ming{-}Wei Chang.
  2020.
\newblock \href {http://arxiv.org/abs/2002.08909} {{REALM:} retrieval-augmented
  language model pre-training}.
\newblock \emph{CoRR}, abs/2002.08909.

\bibitem[{Hamilton et~al.(2017)Hamilton, Ying, and
  Leskovec}]{hamilton2017inductive}
Will Hamilton, Zhitao Ying, and Jure Leskovec. 2017.
\newblock Inductive representation learning on large graphs.
\newblock \emph{Advances in neural information processing systems}, 30.

\bibitem[{Huang et~al.(2020)Huang, Sharma, Sun, Xia, Zhang, Pronin,
  Padmanabhan, Ottaviano, and Yang}]{huang2020embedding}
Jui-Ting Huang, Ashish Sharma, Shuying Sun, Li~Xia, David Zhang, Philip Pronin,
  Janani Padmanabhan, Giuseppe Ottaviano, and Linjun Yang. 2020.
\newblock Embedding-based retrieval in facebook search.
\newblock In \emph{Proceedings of the 26th ACM SIGKDD International Conference
  on Knowledge Discovery \& Data Mining}, pages 2553--2561.

\bibitem[{Kipf and Welling(2016)}]{kipf2016semi}
Thomas~N Kipf and Max Welling. 2016.
\newblock Semi-supervised classification with graph convolutional networks.
\newblock \emph{arXiv preprint arXiv:1609.02907}.

\bibitem[{Lewis et~al.(2021)Lewis, Stenetorp, and Riedel}]{lewis2021question}
Patrick Lewis, Pontus Stenetorp, and Sebastian Riedel. 2021.
\newblock Question and answer test-train overlap in open-domain question
  answering datasets.
\newblock In \emph{Proceedings of the 16th Conference of the European Chapter
  of the Association for Computational Linguistics: Main Volume}, pages
  1000--1008.

\bibitem[{Li et~al.(2021)Li, Lv, Jin, Lin, Yang, Zeng, Wu, and
  Ma}]{li2021embedding}
Sen Li, Fuyu Lv, Taiwei Jin, Guli Lin, Keping Yang, Xiaoyi Zeng, Xiao-Ming Wu,
  and Qianli Ma. 2021.
\newblock Embedding-based product retrieval in taobao search.
\newblock In \emph{Proceedings of the 27th ACM SIGKDD Conference on Knowledge
  Discovery \& Data Mining}, pages 3181--3189.

\bibitem[{Li et~al.(2016)Li, Zemel, Brockschmidt, and Tarlow}]{li2016gated}
Yujia Li, Richard Zemel, Marc Brockschmidt, and Daniel Tarlow. 2016.
\newblock Gated graph sequence neural networks.
\newblock In \emph{Proceedings of ICLR'16}.

\bibitem[{Liu et~al.(2019)Liu, Ott, Goyal, Du, Joshi, Chen, Levy, Lewis,
  Zettlemoyer, and Stoyanov}]{liu2019roberta}
Yinhan Liu, Myle Ott, Naman Goyal, Jingfei Du, Mandar Joshi, Danqi Chen, Omer
  Levy, Mike Lewis, Luke Zettlemoyer, and Veselin Stoyanov. 2019.
\newblock Roberta: A robustly optimized bert pretraining approach.
\newblock \emph{arXiv preprint arXiv:1907.11692}.

\bibitem[{Mao et~al.(2021)Mao, He, Liu, Shen, Gao, Han, and
  Chen}]{mao-etal-2021-generation}
Yuning Mao, Pengcheng He, Xiaodong Liu, Yelong Shen, Jianfeng Gao, Jiawei Han,
  and Weizhu Chen. 2021.
\newblock \href {https://doi.org/10.18653/v1/2021.acl-long.316}
  {Generation-augmented retrieval for open-domain question answering}.
\newblock In \emph{Proceedings of the 59th Annual Meeting of the Association
  for Computational Linguistics and the 11th International Joint Conference on
  Natural Language Processing (Volume 1: Long Papers)}, pages 4089--4100,
  Online. Association for Computational Linguistics.

\bibitem[{Mikolov et~al.(2013)Mikolov, Sutskever, Chen, Corrado, and
  Dean}]{mikolov2013distributed}
Tomas Mikolov, Ilya Sutskever, Kai Chen, Greg~S Corrado, and Jeff Dean. 2013.
\newblock Distributed representations of words and phrases and their
  compositionality.
\newblock \emph{Advances in neural information processing systems}, 26.

\bibitem[{Nigam et~al.(2019)Nigam, Song, Mohan, Lakshman, Ding, Shingavi, Teo,
  Gu, and Yin}]{nigam2019semantic}
Priyanka Nigam, Yiwei Song, Vijai Mohan, Vihan Lakshman, Weitian Ding, Ankit
  Shingavi, Choon~Hui Teo, Hao Gu, and Bing Yin. 2019.
\newblock Semantic product search.
\newblock In \emph{Proceedings of the 25th ACM SIGKDD International Conference
  on Knowledge Discovery \& Data Mining}, pages 2876--2885.

\bibitem[{O{\u{g}}uz et~al.(2021)O{\u{g}}uz, Lakhotia, Gupta, Lewis, Karpukhin,
  Piktus, Chen, Riedel, Yih, Gupta et~al.}]{ouguz2021domain}
Barlas O{\u{g}}uz, Kushal Lakhotia, Anchit Gupta, Patrick Lewis, Vladimir
  Karpukhin, Aleksandra Piktus, Xilun Chen, Sebastian Riedel, Wen-tau Yih,
  Sonal Gupta, et~al. 2021.
\newblock Domain-matched pre-training tasks for dense retrieval.
\newblock \emph{arXiv preprint arXiv:2107.13602}.

\bibitem[{Sachan et~al.(2021)Sachan, Patwary, Shoeybi, Kant, Ping, Hamilton,
  and Catanzaro}]{sachan-etal-2021-end}
Devendra Sachan, Mostofa Patwary, Mohammad Shoeybi, Neel Kant, Wei Ping,
  William~L. Hamilton, and Bryan Catanzaro. 2021.
\newblock \href {https://doi.org/10.18653/v1/2021.acl-long.519} {End-to-end
  training of neural retrievers for open-domain question answering}.
\newblock In \emph{Proceedings of the 59th Annual Meeting of the Association
  for Computational Linguistics and the 11th International Joint Conference on
  Natural Language Processing (Volume 1: Long Papers)}, pages 6648--6662,
  Online. Association for Computational Linguistics.

\bibitem[{Zhang et~al.(2020)Zhang, Wang, Zhang, Tang, Jiang, Xiao, Yan, and
  Yang}]{zhang2020towards}
Han Zhang, Songlin Wang, Kang Zhang, Zhiling Tang, Yunjiang Jiang, Yun Xiao,
  Weipeng Yan, and Wen-Yun Yang. 2020.
\newblock Towards personalized and semantic retrieval: An end-to-end solution
  for e-commerce search via embedding learning.
\newblock In \emph{Proceedings of the 43rd International ACM SIGIR Conference
  on Research and Development in Information Retrieval}, pages 2407--2416.

\end{thebibliography}
\bibliographystyle{acl_natbib}

\appendix
\section{Appendix: Additional nDCG scores of product search performance}
\label{apd:ndcg}
Extending the results of Table \ref{tab: product_search_results}, this section showed three more k values to the nDCG performance of fine-tuned models on product search tasks. Table \ref{tab: ndcg@1} shows nDCG@1, Table \ref{tab: ndcg@20} shows nDCG@20, and Table \ref{tab: ndcg@100} shows nDCG@100. 

The results from nDCG@20 and nDCG@100 are consistent with nDCG@50, such that the qq model achieves the best performance on both seen and unseen (query, product) pair segments. However, in nDCG@1 the Roberta-large model performs the best. This reveals that information gained from domain-specific pre-training makes it more difficult for the model to optimize the highest relevance score range; however, it helps the model to produce a generally better ranked list. Since semantic search models are usually used as candidate generators and followed by dedicated ranking models, the nDCG@1 performance is not as important as nDCG@50 or nDCG@100, and thus we still think domain-specific pre-training is a suitable approach to improve encoders' performance.

\begin{table*}[t]
    \centering
        \begin{tabular}{l||c|c|c|cccc}
        \hline\hline
        &&&&\multicolumn{4}{c}{Unseen sub-buckets} \\
        Bucket name           & Overall   & Seen      & Unseen    & q+, p+    & q+, p-    & q-, p+    & q-, p-\\ \hline
        \hline\hline
        Roberta base                    &0.7845	&0.8858	&0.7793	&\textbf{0.7988}	&0.8061	&0.7204	&0.7446\\
        Roberta large                   &\textbf{0.7969}	&\textbf{0.9032}	&\textbf{0.7892}	&0.7811	&0.807	&0.7231	&0.746\\
        SimCSE large                    &0.7876	&0.8829	&0.7793	&0.7862	&0.8019	&0.7124	&0.7554\\
        Query-query (qq) pretrain       &0.7919	&\textbf{0.9032}	&0.7816	&0.7736	&\textbf{0.8123}	&\textbf{0.746}	&\textbf{0.7742}\\
        Product-product (pp) pretrain   &0.7949	&0.9017	&0.7886	&0.787	&0.812	&0.7272	&0.7433\\
        qq+pp pretrain                  &0.7764	&0.8916	&0.7712	&0.771	&\textbf{0.8123}	&0.7164	&0.7392\\
        \hline\hline
        \end{tabular}
        \caption{nDCG@1 of fine-tuned models on product search tasks. Please refer to Table \ref{tab: product_search_results} for notation definitions.}
    \label{tab: ndcg@1}
\end{table*}

\begin{table*}[t]
    \centering
        \begin{tabular}{l||c|c|c|cccc}
        \hline\hline
        &&&&\multicolumn{4}{c}{Unseen sub-buckets} \\
        Bucket name           & Overall   & Seen      & Unseen    & q+, p+    & q+, p-    & q-, p+    & q-, p-\\ \hline
        \hline\hline
        Roberta base                    &0.7884 &0.921  &0.7807 &\textbf{0.7686} &\textbf{0.8053} &0.7012 &0.7269\\
        Roberta large                   &0.7888 &0.9203 &0.7809 &0.7641 &0.8023 &0.7051 &0.7216\\
        SimCSE large                    &0.7818 &0.9183 &0.7739 &0.7631 &0.7978 &0.6941 &0.7223\\
        Query-query (qq) pretrain       &\textbf{0.7914} &\textbf{0.9229} &\textbf{0.7829} &0.7607 &0.8009 &\textbf{0.7165} &\textbf{0.7316}\\
        Product-product (pp) pretrain   &0.7863 &0.9221 &0.7771 &0.7598 &0.7988 &0.7072 &0.7288\\
        qq+pp pretrain                  &0.7877 &0.9188 &0.78   &0.7628 &0.8007 &0.7151 &0.7298\\
        \hline\hline
        \end{tabular}
        \caption{nDCG@20 of fine-tuned models on product search tasks. Please refer to Table \ref{tab: product_search_results} for notation definitions.}
    \label{tab: ndcg@20}
\end{table*}

\begin{table*}[t]
    \centering
        \begin{tabular}{l||c|c|c|cccc}
        \hline\hline
        &&&&\multicolumn{4}{c}{Unseen sub-buckets} \\
        Bucket name           & Overall   & Seen      & Unseen    & q+, p+    & q+, p-    & q-, p+    & q-, p-\\ \hline
        \hline\hline
        Roberta base                    &0.7681 &0.9386 &0.7593 &\textbf{0.7603} &0.7887 &0.7138 &0.7321\\
        Roberta large                   &0.7678 &0.9392 &0.7583 &0.7539 &0.7842 &0.7102 &0.7284\\
        SimCSE large                    &0.7643 &0.9372 &0.7545 &0.7525 &0.7834 &0.7059 &0.7279\\
        Query-query (qq) pretrain       &\textbf{0.7721} &\textbf{0.9407} &\textbf{0.7624} &0.7523 &\textbf{0.7913} &\textbf{0.7229} &\textbf{0.7334}\\
        Product-product (pp) pretrain   &0.7664 &0.9406 &0.7571 &0.7529 &0.7846 &0.7157 &0.7308\\
        qq+pp pretrain                  &0.7703 &0.9377 &0.7609 &0.7539 &0.789 &0.7199 &0.7315\\
        \hline\hline
        \end{tabular}
        \caption{nDCG@100 of fine-tuned models on product search tasks. Please refer to Table \ref{tab: product_search_results} for notation definitions.}
    \label{tab: ndcg@100}
\end{table*}

\end{document}